# Nondestructive quantitative estimation of cross-sectional corrosion degree of rebar using self-magnetic flux leakage field variation

Junli Qiu [a, b], Weiping Zhang [b, c, *], Chao Jiang [b, c], Hong Zhang [a], Jianting Zhou [a]

[a] State Key Laboratory of Mountain Bridge and Tunnel Engineering, Chongqing Jiaotong University, Chongqing 400074, China

[b] Key Laboratory of Performance Evolution and Control for Engineering Structures of Ministry of Education, Tongji University, 1239 Siping Road, Shanghai 200092, China

[c] Department of Structural Engineering, College of Civil Engineering, Tongji University, 1239 Siping Road, Shanghai 200092, China

## Abstract

To accurately assess the structural performance of corroded reinforced concrete structures, obtaining precise information on the corrosion range and corrosion degree of rebar is crucial. In this study, based on a comprehensive analysis of extensive high-precision magnetic field and three-dimensional structural light scanning data of 21 corroded rebars, it was found that the self-magnetic flux leakage can accurately identify the corrosion range, with an error not exceeding 3%. A proposed quantitative index $NH_x$ of the self-magnetic flux leakage amplitude exhibits a linear correlation with the cross-sectional corrosion degree of rebar, whose probability density distribution can be accurately described using the Weibull distribution function. Utilizing the Weibull distribution function of $NH_x$ and a Bayesian model, automatically rapid quantification of the rebar's cross-sectional corrosion degree based on the non-destructive testing-derived $NH_x$ values can be conveniently realized. This self-magnetic flux leakage-based novel method for quantifying rebar's cross-sectional corrosion degree is accurate, efficient, and well-suited for practical engineering applications, providing robust support for a precise assessment of the structural performance of corroded reinforced concrete structures.

| **Nomenclature** | | | |
|---|---|---|---|
| $C_L$ | center X-coordinate of the left intersection areas of the $H_{Sx}$ curves | $NH_{x-m}$ | corrected $NH_x$ |
| $C_R$ | center X-coordinate of the right intersection areas of the $H_{Sx}$ curves | PDD | probability density distribution |
| CR | correction factor of the index $NH_x$ | PDF | probability density function |
| $d$ | corrosion depth | $R^2$ | coefficient of determination |
| $f(\cdot)$ | Function | RC | Reinforced concrete |
| $f(\cdot|\cdot)$ | likelihood function | RE | relative error |
| $H$ | total magnetic field strength | $r$ | radius of the rebar |
| $H_x$ | tangential component of $H$ | $S_0$ | cross-sectional area of the uncorroded rebar |
| $H_z$ | normal component of $H$ | $S_c$ | corroded rebar's cross-sectional area |
| $H_I$ | induced magnetic field strength | SMFL | self-magnetic flux leakage |
| $H_{I-av}$ | average value of the modulus of $H_I$ | $t$ | corrosion duration |
| $H_{Ix}$ | tangential component of $H_I$ | $W_m$ | measured corrosion width |
| $H_{Iz}$ | normal component of $H_I$ | $w$ | half-corrosion width |
| $H_S$ | SMFL field strength | $x$ | X-coordinate variable |
| $H_{Sx}$ | tangential component of $H_S$ | $y$ | Y-coordinate variable |
| $H_{Sz}$ | normal component of $H_S$ | $z$ | Z-coordinate variable |
| $I$ | symbol of quantitative SMFL index | $\eta$ | rebar's cross-sectional corrosion degree |
| $i_{corr}$ | corrosion current density | $\lambda$ | scale parameter of the Weibull distribution function |
| KDE | kernel density estimation | $\pi(\cdot)$ | prior distribution |
| $k$ | shape parameter of the Weibull distribution function | $\pi(\cdot|\cdot)$ | posterior distribution |
| LFH | Lift-off height | $\rho_s$ | density of iron |
| $l$ | half-length of rebar | $\Theta$ | parameter space |
| $NH_x$ | a quantitative SMFL index | | |

# 1. Introduction

Reinforced concrete (RC) structures are extensively utilized in infrastructure projects, including bridges, buildings, and tunnels [1-3]. Under the sustained actions of various loads and aggressive environmental conditions, the rebars within RC structures will experience corrosion, leading to a reduction in the cross-sectional area of the rebars, degradation of the bond behavior between rebars and concrete, and cracking or even delamination of the concrete cover [4-6]. Consequently, these issues result in a deterioration of the structural load-bearing capacity [7-9], a decline in overall structural reliability [10-11], and even premature structural failure, which can cause significant loss of casualties and damage to societal property. Therefore, timely detection and accurate assessment of the initialization and propagation of rebar corrosion are essential for ensuring the long-term safety of RC structures.

Influenced by factors such as relative humidity, oxygen content, and material properties, the corrosion process of rebars in RC structures is characterized by high stochastic variability [12-14]. The rebar's corrosion is typically detected using both electrochemical and physical techniques. While electrochemical techniques, such as half-cell potential measurements [15-16], can offer valuable insights into the true corrosion state or rate of the rebars, they often encounter difficulties in accurately quantifying the corrosion degree. Physical methods, including acoustic emission [17], fiber optic sensing [18], and infrared thermography [19], can qualitatively estimate the degree of rebar corrosion; however, they are susceptible to the effects of random environmental conditions, material properties, and the complex parameters of the corrosion system, complicating efforts to quantify the corrosion degree. Although X/γ-ray computed tomography [20] facilitates precise imaging and high-resolution detection of rebar corrosion, challenges such as radiation contamination, high costs, and low efficiency restrict its practical applications in engineering. In contrast, electromagnetic detection techniques like eddy current [21], ground-penetrating radar [22], and self-magnetic flux leakage (SMFL) [23] may still exhibit gaps in evaluation precision but are generally more suitable for engineering applications. Among these, SMFL detection is especially favored due to its low cost, high efficiency, and strong resistance to interference, establishing it as the preferred technique for detecting rebar corrosion.

Rebars are composed of ferromagnetic materials and can generate a magnetic field under various factors, including rolling stress and external magnetic fields [4, 24-25]. When corrosion defects manifest in the rebars, the magnetic field lines within the rebar will spontaneously leak to the exterior, leading to changes in the magnetic field surrounding the corrosion defects; this phenomenon is referred to as SMFL field variation [4, 24-25]. On the one hand, SMFL field variation can effectively indicate both the location and degree of rebar corrosion [26, 27]. On the other hand, the amplitude of the SMFL field variation can be utilized to assess the corrosion degree of the rebars. Many robust linear relationship models between the corrosion degree and the amplitude of the SMFL field variation of rebars have been established [4, 28-32]. However, due to significant adverse interference from the random magnetization of the rebars, a considerable magnitude difference exists in the slopes of these linear models. Consequently, it is not feasible to estimate the rebar corrosion degree based on the amplitude of the SMFL field variation when the magnetization strength remains unknown. To address this challenge, our research group has developed a quantification method for estimating rebar corrosion degrees that effectively avoids the adverse effects of random magnetization, grounded in the principles of ferromagnetism and magnetic dipole theory [32]. By employing this quantification method, our group has successfully achieved a non-destructive probabilistic quantitative estimation of the maximum cross-sectional corrosion degree of locally pitted corroded rebars based on SMFL [23, 33-34]. However, in reinforced concrete structures, in addition to localized pitting corrosion, the entire corrosion of rebars may also occur [5], and, as of now, an SMFL-based quantification estimation for the cross-sectional corrosion degree of entirely corroded rebar has not been realized for this scenario. Thus, a significant research gap remains that requires further investigation.

In this study, three-dimensional (3D) structured light and SMFL scanning were initially employed to gather extensive data on the cross-sectional corrosion degrees and the corresponding SMFL field variations of entirely corroded rebars. Subsequently, the correlation between the cross-sectional corrosion degree of the rebars and the quantitative SMFL index $NH_x$ was analyzed, leading to the development of a correlation model from a probabilistic perspective. Finally, based on this correlation model and Bayesian theory, the non-destructive quantification estimation of the cross-sectional corrosion degree of entirely corroded rebars was successfully realized using SMFL.

## 2. The experimental details

## 2.1. The corrosion of rebar specimens

In RC structures, the rebars are embedded in the concrete, and corrosion products accumulate on its surface once corrosion initiates. However, concrete and corrosion products have been demonstrated to affect magnetic field detection almost identically to the air [5, 26, 29-31, 35]. Therefore, to delve into the inherent relationship between the corrosion degree and the SMFL of the entirely corroded rebar, this study ignores the minor interference with the concrete and the corrosion products.

The experiments were conducted using 21 bare hot-rolled threaded rebar specimens, each measuring 1500 mm in length and 14 mm in diameter, labeled S1 to S21. The impressed current method [2, 4] was employed to simulate the entire corrosion in the central 1000 mm segment of the rebar specimens, as shown in Fig. 1. Generally, the impressed current density during accelerated corrosion does not exceed 200 μA/cm² [36-38]. To explore the potential effects of different current densities on SMFL, the specimens were equally divided into three groups, with current densities of 200 μA/cm², 600 μA/cm², and 1200 μA/cm², respectively.

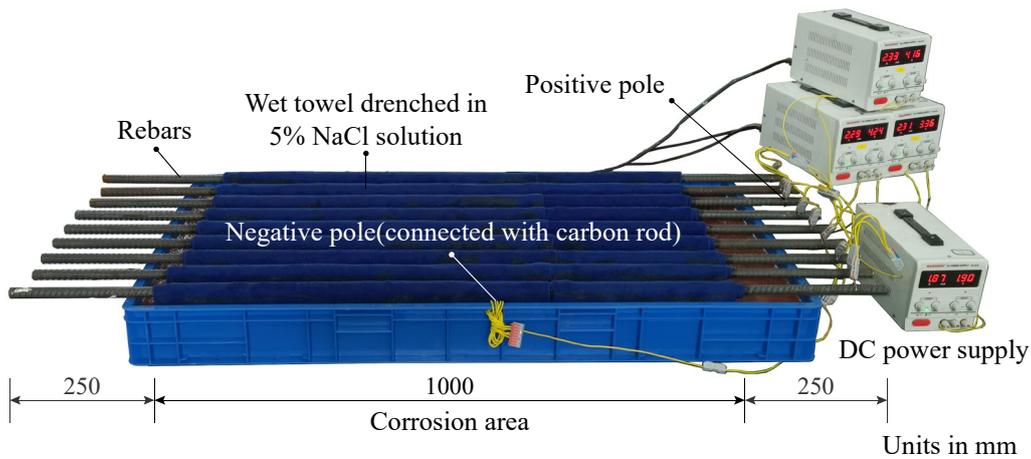

**Fig. 1.** Applied current corrosion of rebar specimens S1~S21.

The relationship between the theoretical corrosion degree $\eta_t$ of uniform corroded rebar and the corrosion duration $t$ is given by the following equation:

$$t \approx \frac{2 \cdot 96487 \cdot \rho_s \cdot r \cdot \eta_t}{i_{corr} \cdot 55.8} \qquad (1)$$

where 2 is the valence state of $Fe^{2+}$, 96487 is Faraday's constant, $\rho_s = 7.86$ g·cm$^{-3}$ is the density of iron, $r$ is the radius of the rebar, $i_{corr}$ is the corrosion current density, and 55.8 is the relative atomic mass of iron.

The $\eta_t$ for all rebar specimens was set at 0.2. As the actual radius $r$ of the rebar specimens decreases during corrosion, the corrosion surface area gradually reduces. Under constant impressed current, the $i_{corr}$ tends to increase; thus, the corrosion duration $t$ calculated according to Eq. (1) and the radius of uncorroded rebar is an approximation. The $i_{corr}$ and corrosion durations $t$ for the three sets of rebar specimens are presented in Table 1.

Table1 The corrosion current densities $i_{corr}$ and corrosion durations $t$ for the three sets of rebar specimens.

| No. | $i_{corr}$ (μA/cm$^2$) | $\eta_t$ | $t$ (h) |
|---|---|---|---|
| S1~S7 | 200 | 0.2 | 2654.4 |
| S8~S14 | 600 | 0.2 | 884.8 |
| S15~S21 | 1200 | 0.2 | 442.4 |

## 2.2. The SMFL field scanning of the corroded rebar specimens

Before conducting the impressed current corrosion, longitudinal magnetic field scans were performed on all the rebar specimens to obtain the induced magnetic field $H_I$ information of the rebar specimens in their uncorroded state, as shown in Fig. 2. The vertical distance between the magnetic sensors and the upper surface of the rebar is defined as the Lift-off height (LFH). Each rebar specimen was scanned six times with different LFH values of 5, 15, 30, 50, 70, and 100 mm. After completing the impressed current corrosion, longitudinal magnetic field scans were conducted again along these six paths to obtain information on the SMFL field $H_S$ variations of the corroded rebar specimens.

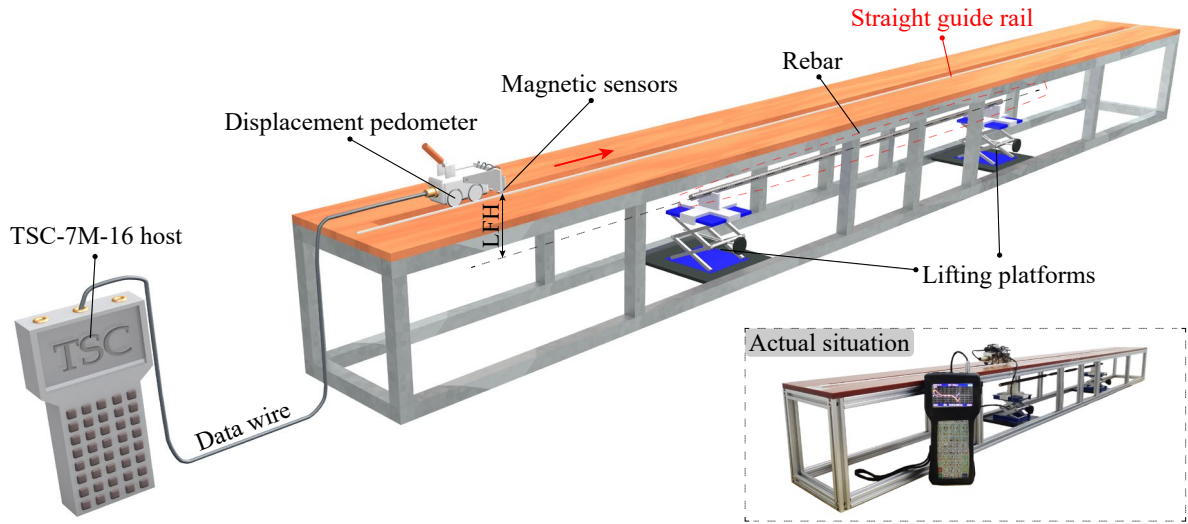

**Fig. 2.** Schematic diagram of SMFL field scanning of corroded rebars.

## 2.3 The 3D structured light scanning of corroded rebar specimens

To obtain the actual cross-sectional corrosion degree of the corroded rebar specimen, a three-dimensional structured light scanning technique, as shown in Fig. 3, was employed to scan the corroded rebars and acquire their precise geometric model. Initially, a structured light projector projects blue light with specific patterns onto the surface of the corroded rebar, and the reflected light is captured by a camera. By utilizing the principles of triangulation, camera modeling, and the geometric characteristics of the light path, the spatial position of any point on the rebar surface can be calculated [5, 39]. Surface scanning allows for the acquisition of point cloud data of the corroded rebar surface, which, after wrapped processing, results in a three-dimensional geometric model of the corroded rebar specimen with an accuracy of up to 0.02 mm.

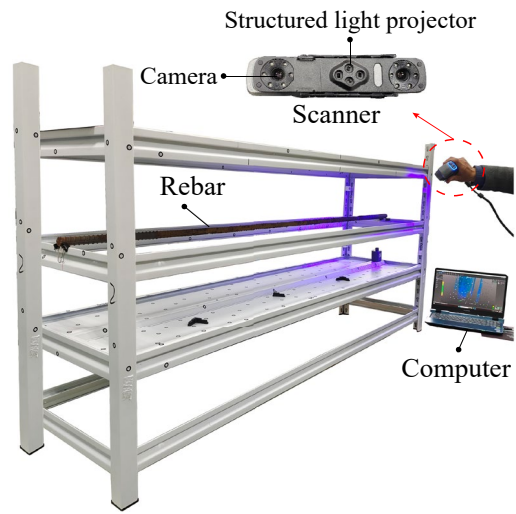

Fig. 3. The 3D structured light scanning of corroded rebar specimens.

## 3. Results and discussions

### 3.1. Geometric models and magnetic field variations of corroded rebars

As illustrated in Fig. 4, taking the accurate geometric model of the corroded rebar specimen S1 as an example, it is observable that the cross-section of the rebar undergoes an overall reduction after corrosion, accompanied by the development of uneven corrosion morphology. Utilizing this accurate geometric model, the three-dimensional measurement software facilitates the extraction of accurate cross-sectional profiles of the rebar specimen S1 at various X-coordinate values using the cross-section slices, some of which are depicted in Fig. 4. In the uncorroded area of X = 50 mm and X = 1450 mm, the cross-sections of rebar specimen S1 distinctly reveal the presence of ribs. Conversely, in the corroded area, such as at X = 350 mm, the cross-sectional profiles of the rebar specimen S1 are markedly smaller, and the ribs are no longer easily discernible. These cross-sectional profiles enable the computation of corroded rebar's cross-sectional area $S_c$ values with an accuracy of 0.01 mm² using the three-dimensional measurement software.

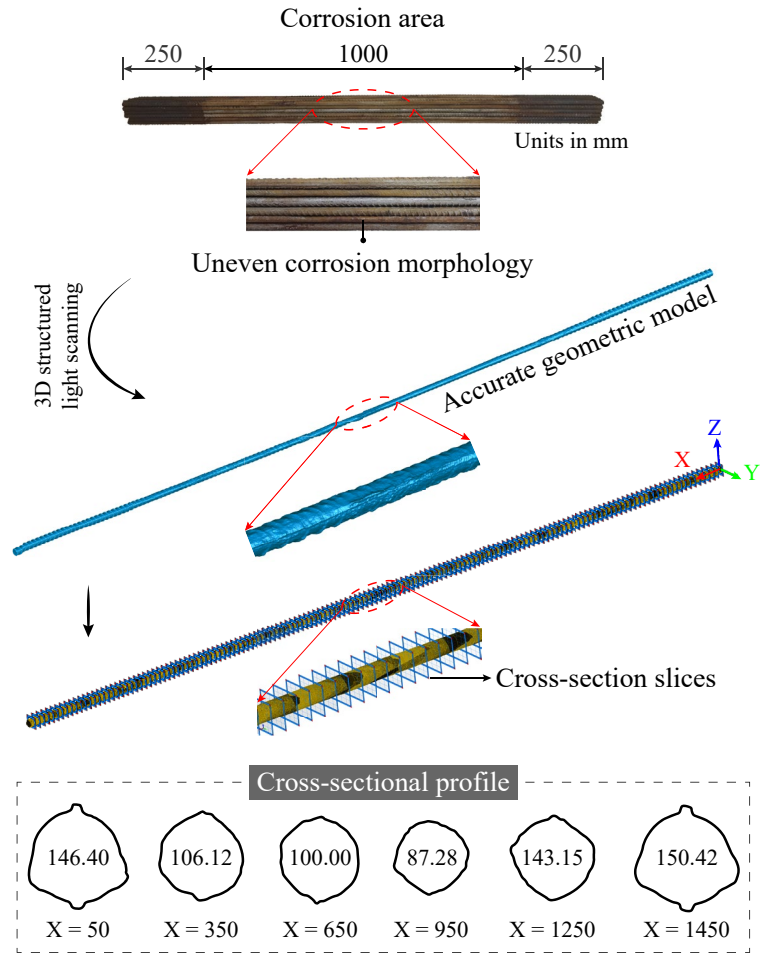

**Fig. 4.** The accurate geometric model and typical cross-sectional profiles of the rebar specimen S1

The characteristics of the corrosion-induced SMFL field variations in corroded rebar can be comprehensively analyzed using both experimental and simulation data. The simulation data were obtained based on the geometric solid model of the corroded rebar specimen and finite element simulation conducted by COMSOLR software (COMSOL Inc, Stockholm, Sweden), following the same simulation process described in reference [5]. For the sake of brevity, rebar specimen S1 is used as an example to illustrate the SMFL field variations caused by rebar corrosion. As shown in Fig. 5, the curves of the tangential component $H_{Ix}$ and the normal component $H_{Iz}$ of the induced magnetic field $\boldsymbol{H}_I$ with different LFH values of specimen S1 in its uncorroded state are relatively smooth and exhibit good positive or negative symmetry, which aligns well with the simulation results. In contrast, the curves of the tangential component $H_x$ and the normal component $H_z$ of the total magnetic field strength $H$ with different LFH values of specimen S1 in its corroded state show significant fluctuations within the corrosion area of 250 - 1250 mm. The characteristics of these fluctuations are in complete agreement with the simulation results, although there are some differences in magnitude. The total magnetic field strength $H$ comprises the induced magnetic field $H_I$ and the SMFL field $H_S$, and the fluctuations in their component curves are attributed to the corrosion-induced SMFL field variations. Therefore, the corrosion area of the rebar can be accurately located based on the anomalous fluctuations in the SMFL field curves.

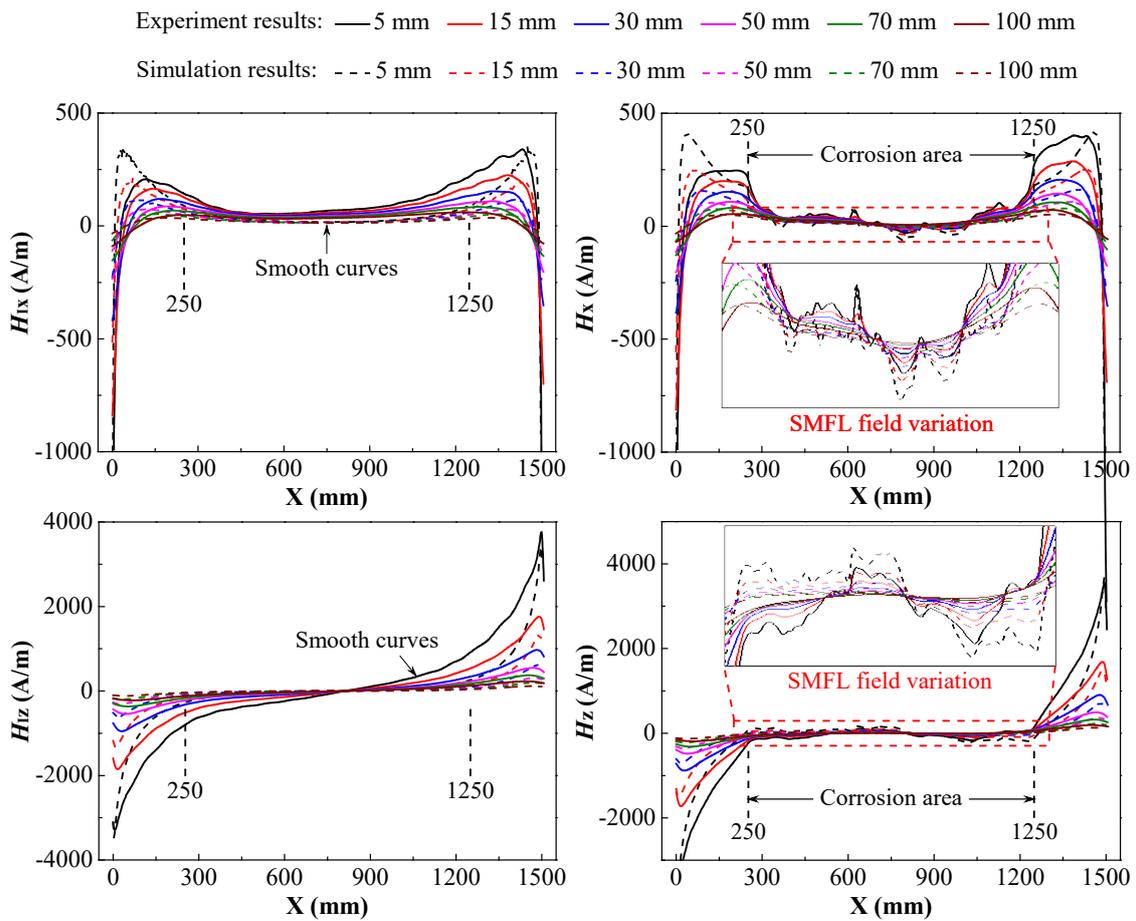

**Fig. 5.** The curves of $H_{Ix}$, $H_{Iz}$, $H_x$, and $H_z$ components with different LFH values of rebar specimen S1.

## 3.2. Correlation between SMFL field variation and cross-sectional area of rebar

Based on Figs. 4 and 5, the correlation between the SMFL field variation and the cross-sectional area of corroded rebar can be further analyzed. The tangential component $H_{Sx}$ and the normal component $H_{Sz}$ of the corrosion-induced SMFL field $\boldsymbol{H}_S$ are obtained by subtracting the $H_{Ix}$ and $H_{Iz}$ values from the $H_x$ and $H_z$ values one-to-one [4, 33], as expressed:

$$\begin{cases} H_{Sx}(i) = H_x(i) - H_{Ix}(i) \\ H_{Sz}(i) = H_z(i) - H_{Iz}(i) \end{cases} \quad (2)$$

where $i$ represents the longitudinal (X) coordinate value of the rebar.

Taking the $H_{Sx}$ curves and the cross-sectional area $S_c$ curve of specimen S1 as an example for correlation analysis. As shown in Fig. 6, the $H_{Sx}$ curves with different LFH values and the $S_c$ curve of specimen S1 are highly consistent in shape, especially when LFH is small. This result exhibits a strong positive correlation between the corrosion-induced cross-sectional area loss and SMFL field variation of rebar. In addition, the $H_{Sx}$ curves exhibit two intersection areas at both ends of the corroded area. Table 2 presents the center X-coordinates $C_L$ and $C_R$ of the left and right intersection areas of the $H_{Sx}$ curves of specimens S1 – S21, showing that $C_L$ and $C_R$ are approximately equal to the actual corrosion boundary X-coordinates of 250 mm and 1250 mm, respectively. Defining the measured corrosion width as $W_m = C_R - C_L$, the calculated $W_m$ values are very close to the designated corrosion width of 1000 mm, with a relative error $RE$ for $W_m$ not exceeding 3.0%. This result indicates that the $H_{Sx}$ curves can be used to estimate the corrosion width accurately.

Table2 Measured corrosion width of all the specimens based on the intersections of the $H_{Sx}$ curves

| No. | $C_L$ | $C_R$ | $W_m$ | RE | No. | $C_L$ | $C_R$ | $W_m$ | RE |
| --- | --- | --- | --- | --- | --- | --- | --- | --- | --- |

|     | (mm) | (mm) | (mm) | (%) |     | (mm) | (mm) | (mm) | (%) |
|-----|------|------|------|-----|-----|------|------|------|-----|
| S1  | 248  | 1246 | 998  | 0.2 | S12 | 257  | 1240 | 983  | 1.7 |
| S2  | 248  | 1255 | 1007 | 0.7 | S13 | 262  | 1232 | 970  | 3.0 |
| S3  | 252  | 1258 | 1006 | 0.6 | S14 | 253  | 1238 | 985  | 1.5 |
| S4  | 250  | 1252 | 1002 | 0.2 | S15 | 266  | 1256 | 990  | 1.0 |
| S5  | 250  | 1255 | 1005 | 0.5 | S16 | 260  | 1234 | 974  | 2.6 |
| S6  | 240  | 1255 | 1015 | 1.5 | S17 | 261  | 1264 | 1003 | 0.3 |
| S7  | 254  | 1256 | 1002 | 0.2 | S18 | 265  | 1249 | 984  | 1.6 |
| S8  | 253  | 1242 | 989  | 1.1 | S19 | 262  | 1254 | 992  | 0.8 |
| S9  | 246  | 1247 | 1001 | 0.1 | S20 | 267  | 1240 | 973  | 2.7 |
| S10 | 265  | 1240 | 975  | 2.5 | S21 | 263  | 1255 | 992  | 0.8 |
| S11 | 260  | 1243 | 983  | 1.7 |     |      |      |      |     |

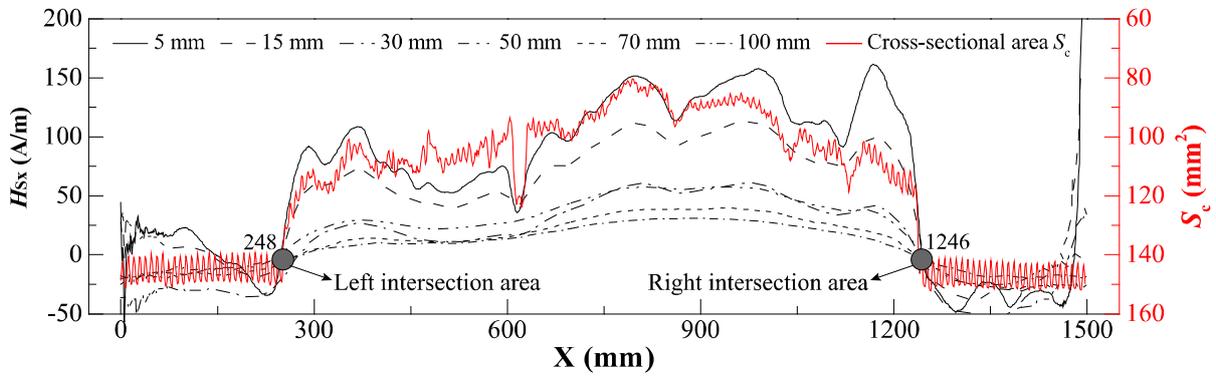

**Fig. 6.** The $H_{Sx}$ curves with different LFHs and cross-sectional area $S_c$ curve of specimen S1.

### 3.3. Correlation between an SMFL quantitative index $NH_x$ and corrosion degree $\eta$

Once the corrosion area is identified using the SMFL field variation, the next crucial step is to estimate the corrosion degree of the rebar non-destructively based on the SMFL field variation. However, the $H_{Sx}$ and $S_c$ curves alone do not sufficiently elucidate the quantitative relationship between the SMFL field variation and the corrosion degree of rebar. By establishing effective quantitative indices related to the cross-sectional area and the SMFL field variation of the corroded rebar, further analysis of their correlation can be performed. Our group has proposed a theory and methodology to construct any quantification SMFL index using the SMFL field variation information of the corroded rebar [32]. Based on this theory, a quantitative SMFL index $NH_x$ can be defined:

$$NH_x(i) = \frac{H_{Sx}(i)}{H_{I\text{-av}}} = \frac{H_{Sx}(i)}{\frac{1}{1000} \cdot \left( \sum_{i=250}^{1250} \sqrt{H_{Ix}(i)^2 + H_{Iz}(i)^2} \right)} \tag{3}$$

where $H_{I\text{-av}}$ is the average value of the modulus of the specimen's $\mathbf{H}_I$ within the corrosion area of 250 mm to 1250 mm, and $i$ represents the longitudinal X-coordinate value of the rebar.

Using the cross-sectional area $S_0$ of the uncorroded rebar specimen and $S_c$ of the corroded rebar specimen, the rebar's cross-sectional corrosion degree $\eta$ is defined as:

$$\eta = \frac{S_0 - S_c}{S_0} \tag{4}$$

As shown in Fig. 7, the $NH_x$ curves for specimen S1 at different LFHs are more clustered compared to the $H_{Sx}$ curves and are also closer in shape to the $\eta$ curve. Additionally, as two dimensionless indices, the curves of $NH_x$ and $\eta$ are quite similar in magnitude, demonstrating a significant positive correlation.

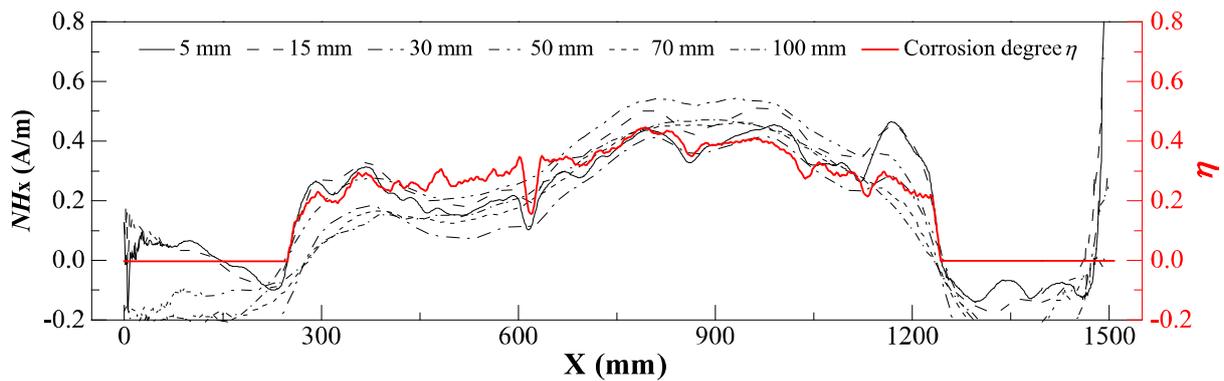

**Fig. 7.** The $NH_x$ curves with different LFHs and corrosion degree $\eta$ curve of specimen S1.

The $H_{Sx}$ and $NH_x$ sample points in the range of 350 mm to 1150 mm of all rebar specimens are presented as scatter plots in Figs. 8(a) and 8(b), respectively, with $\eta$ as the horizontal axis. Considering the notable effects of localized corrosion within a 100 mm range at both ends of the corrosion area, those $H_{Sx}$ and $NH_x$ sample points are excluded. With increasing LFH, the $H_{Sx}$ sample points exhibit distinct partitioning, and no significant correlation with $\eta$ is observed. In contrast, the distribution of $NH_x$ sample points shows a pronounced linear correlation with $\eta$.

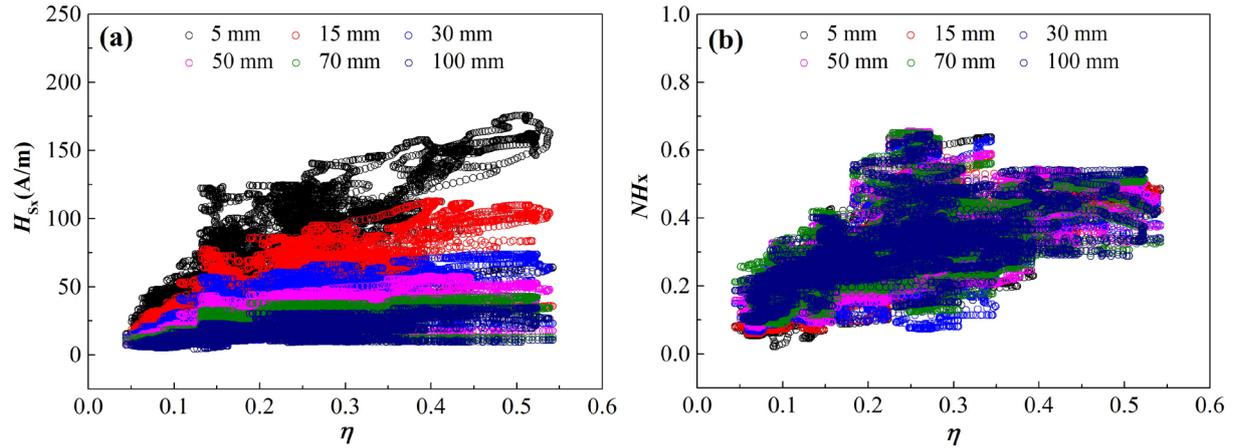

**Fig. 8.** (a) $H_{Sx}$ and (b) $NH_x$ sample points for all rebar specimens with different LFHs.

### 3.4. The probability density distribution (PDD) characteristics of index $NH_x$

In Fig. 8, there are a total of 100,800 sample points, which belong to six LFHs and three corrosion current densities $i_{corr}$. Therefore, the differences in the PDD characteristics of $NH_x$ samples from different LFHs and $i_{corr}$ are analyzed in this section.

#### 3.4.1. Effect of LFH on the PDD of $NH_x$

As shown in Fig. 9, with the increase of corrosion degree $\eta$, the $NH_x$ sample points of different LFHs exhibit an apparent linear growth trend. The slopes of these linear fitting functions range from 1.076 to 1.246 and show an increasing trend with the LFH, and the coefficient of determination $R^2$ ranges from 0.885 to 0.911. The slope approaching 1.0 indicates that once the rebar cross-section suffers from corrosion damage, the internal magnetic flux lines proportionally "leak" to the external, forming an SMFL field. This quantitative relationship compellingly demonstrates the close correlation between the corrosion-induced SMFL field variation and the cross-section loss of rebar, highlighting the significant potential of using SMFL field variation for the non-destructive estimation of rebar corrosion degree.

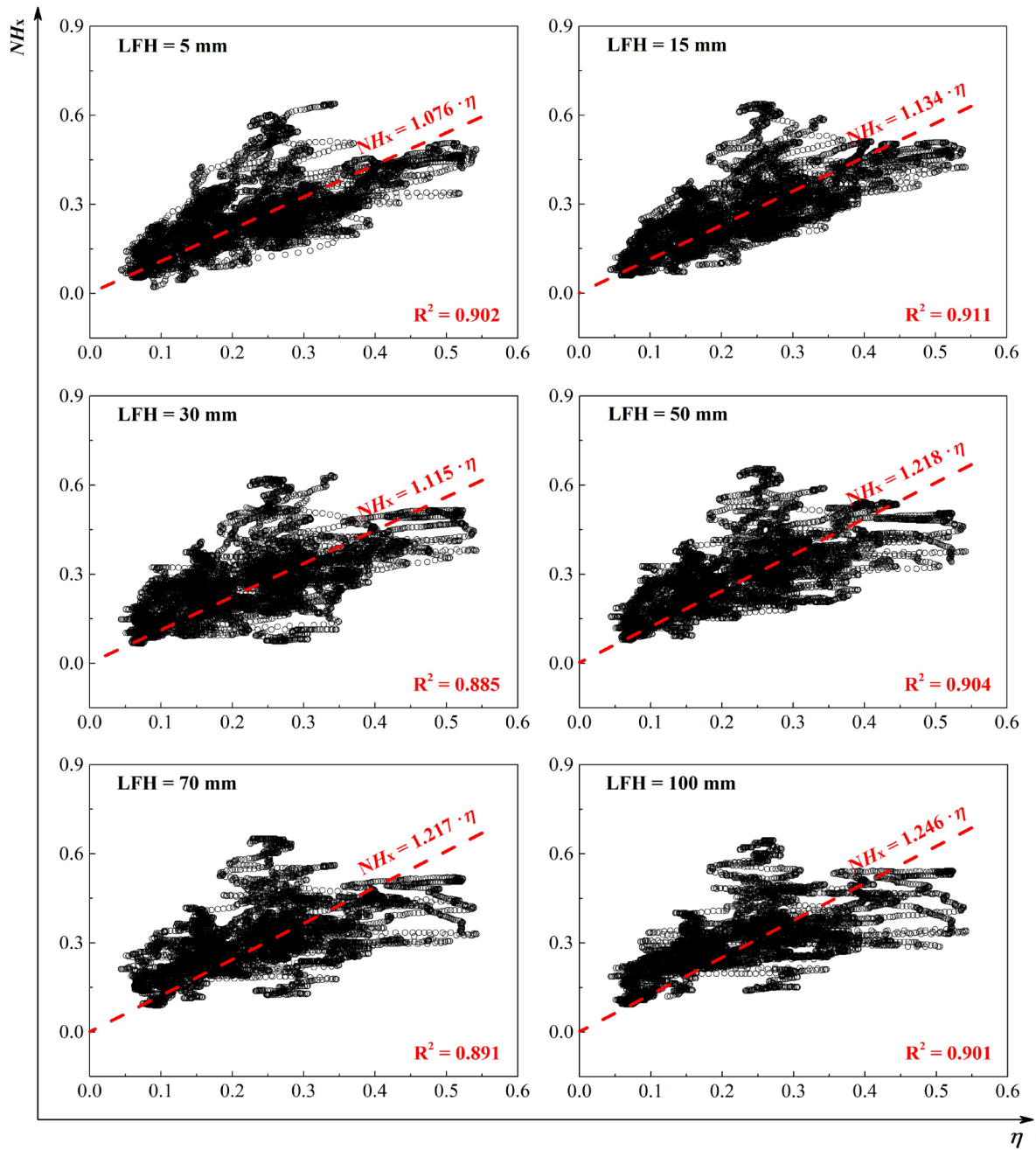

**Fig. 9.** The $NH_x$ sample points with different LFHs.

As shown in Fig. 9, the distribution of $NH_x$ sample points with different LFHs exhibits a high degree of similarity. Therefore, a similarity analysis in the PDD with $NH_x$ sample points of different LFHs is conducted as follows.

**Step 1:** According to the actual distribution of $NH_x$ sample points in Fig. 9, the corrosion degree $\eta$ range of [0.04, 0.55] shown in Fig. 9 is divided into ten sub-intervals of [0.04, 0.08), [0.08, 0.12), [0.12, 0.16), [0.16, 0.20), [0.20, 0.24), [0.24, 0.28), [0.28, 0.32), [0.32, 0.36), [0.36, 0.40), [0.40, 0.55].

**Step 2:** The kernel density estimation (KDE) method [10, 40] for discrete samples is employed to determine the actual PDD of $NH_x$ sample points within the aforementioned ten $\eta$ sub-intervals with six different LFHs.

**Step 3:** Within each $\eta$ sub-interval, there are six actual corresponding PDD sequences of $NH_x$ for six LFHs. All pairwise combinations of these six sequences yield 30 sequence combinations.

**Step 4:** The Pearson correlation coefficient and Jensen-Shannon (JS) divergence for these 30 sequence combinations within each $\eta$ sub-interval are calculated.

The Pearson correlation coefficient is used to evaluate the linear correlation between the two PDD sequences $X$ and $Y$ of $NH_x$ [41]:

$$P_{X,Y} = \frac{\sum_{i=1}^{n}(X_i - \overline{X})(Y_i - \overline{Y})}{\sqrt{\sum_{i=1}^{n}(X_i - \overline{X})^2} \sqrt{\sum_{i=1}^{n}(Y_i - \overline{Y})^2}} \tag{5}$$

where $\overline{X}$ and $\overline{Y}$ represent the sample means of the PDD sequences $X$ and $Y$ of the index $NH_x$.

The transformation of sequences $X$ and $Y$ to $(a + bX)$ and $(c + dY)$, respectively, does not change the value of the Pearson correlation coefficient. Therefore, a higher Pearson correlation coefficient value indicates a stronger linear relationship between sequences $X$ and $Y$ but does not guarantee that their locations and scales are similar. To address this, the JS divergence is introduced [42]:

$$\mathrm{JS}(X\|Y) = \frac{1}{2}\sum_{i=1}^{n} p(x_i) \log \frac{2p(x_i)}{p(x_i)+p(y_i)} + \frac{1}{2}\sum_{i=1}^{n} p(y_i) \log \frac{2p(y_i)}{p(x_i)+p(y_i)} \quad (6)$$

where $p(x_i)$ and $p(y_i)$ are the probability density values of the samples $x_i$ and $y_i$ of the PDD sequences $X$ and $Y$, respectively.

It can be proven that if the distributions of sequences $X$ and $Y$ are completely non-overlapping, the JS divergence is $\log 2 \approx 0.301$, regardless of how close the distribution centers of the two sequences are [43]. Therefore, employing the Pearson correlation coefficient and JS divergence together allows for a comprehensive assessment of the distribution similarity in sequences $X$ and $Y$. In other words, as the Pearson correlation coefficient approaches 1.0 and the JS divergence approaches 0, the distributions of sequences $X$ and $Y$ become increasingly similar in both shape and location. When the Pearson correlation coefficient equals 1.0 and the JS divergence equals 0, sequences $X$ and $Y$ are identical.

The Pearson correlation coefficients and JS divergence matrices for the $NH_x$'s PDD sequence combinations with different LFHs of different $\eta$ sub-intervals are shown in Figs. 10 and 11, where LFH numbers 1, 2, 3, 4, 5, and 6 correspond to LFHs of 5, 15, 30, 50, 70, and 100 mm, respectively. Across all $\eta$ sub-intervals, the Pearson correlation coefficient values generally exceed 0.8, and the JS divergence values under 0.1. The means of the Pearson correlation coefficient and JS divergence of each $\eta$ sub-interval range from 0.76 to 0.88 and from 0.043 to 0.12, respectively, with overall means of 0.844 and 0.045. These results indicate a high degree of similarity in the $NH_x$'s PDDs with different LFHs. The few low Pearson correlation coefficient values and high JS divergence values in the upper left or lower right corners of the matrix suggest that greater differences in LFH lead to lower similarity in the $NH_x$'s PDDs. In practical engineering, magnetic field scans are typically conducted on concrete surfaces, where the concrete layer thickness usually ranges from 30 to 50 mm; thus, LFHs of 5 mm, 15 mm, and 100 mm are generally not encountered. Therefore, when limiting the LFH number range to 3 – 5, the similarity of $NH_x$'s PDDs is higher, which is advantageous for practical engineering applications. Consequently, considering the high similarity of $NH_x$'s PDDs at different LFHs, it is believed that the LFH does not significantly affect the characteristics of $NH_x$'s PDD.

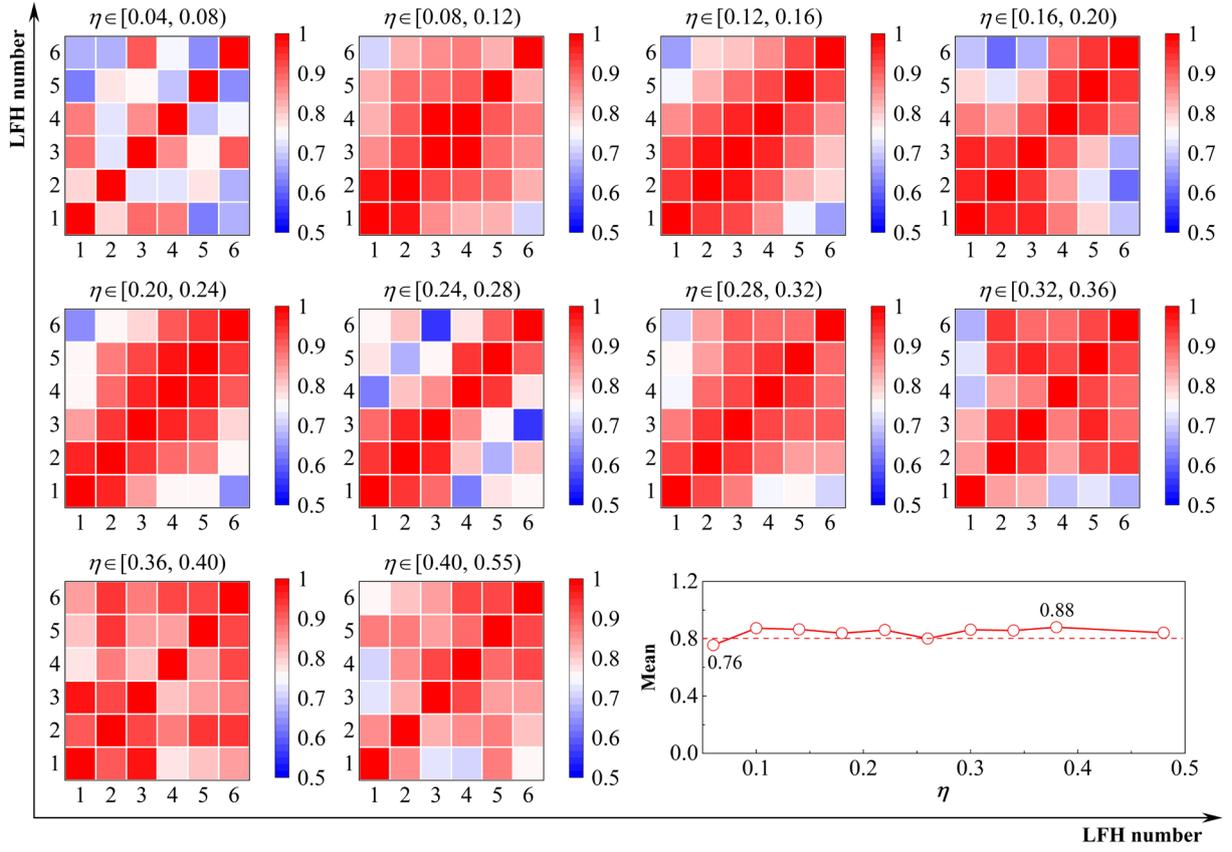

**Fig. 10.** Pearson correlation coefficient of PDD sequence combination of $NH_x$ at different LFHs.

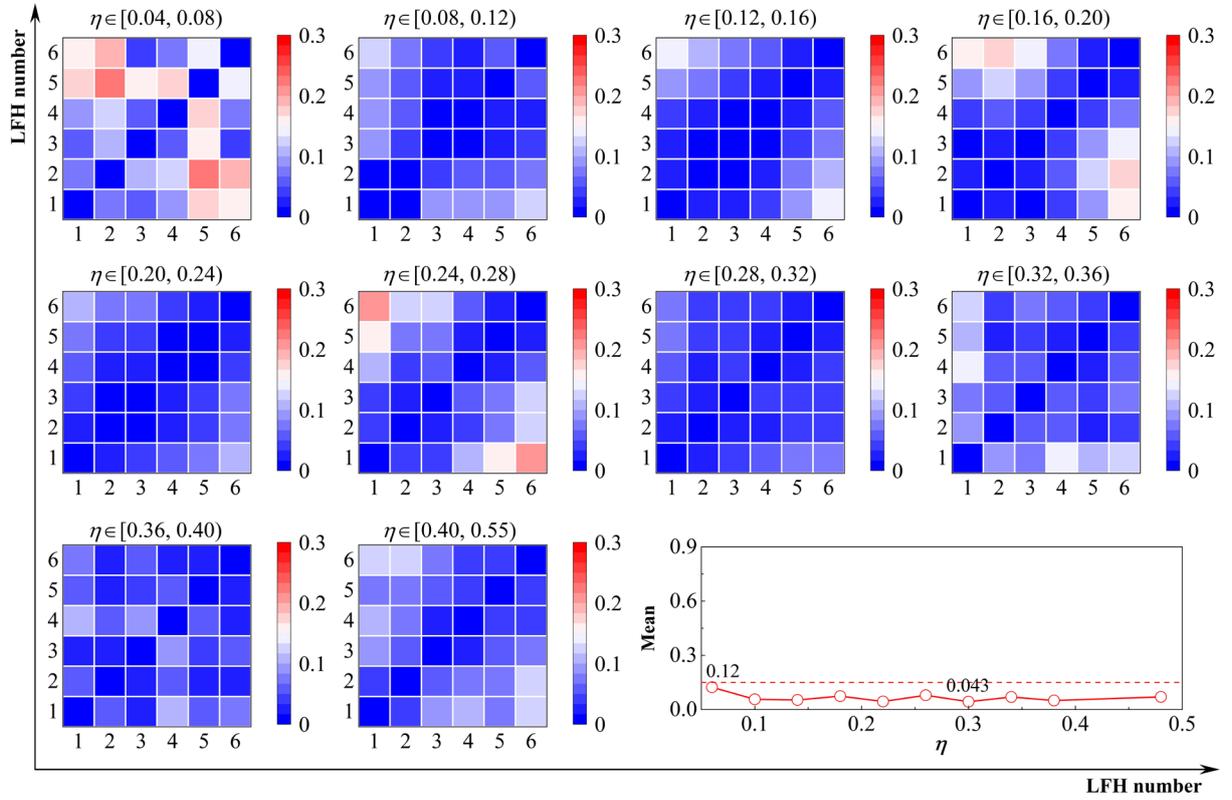

**Fig. 11.** JS divergence of PDD sequence combination of $NH_x$ at different LFHs.

### 3.4.2. Effect of $i_{corr}$ on the PDD of $NH_x$

As shown in Fig. 12, the $NH_x$ sample points with different corrosion current densities $i_{corr}$ exhibit a linear increasing trend with the corrosion degree $\eta$ increases, the slopes of the linear fitting functions range from 1.048 to 1.296, and the coefficient of determination $R^2$ ranges from 0.892 to 0.938. There is no monotonic relationship between the fitting slopes and the $i_{corr}$ observed.

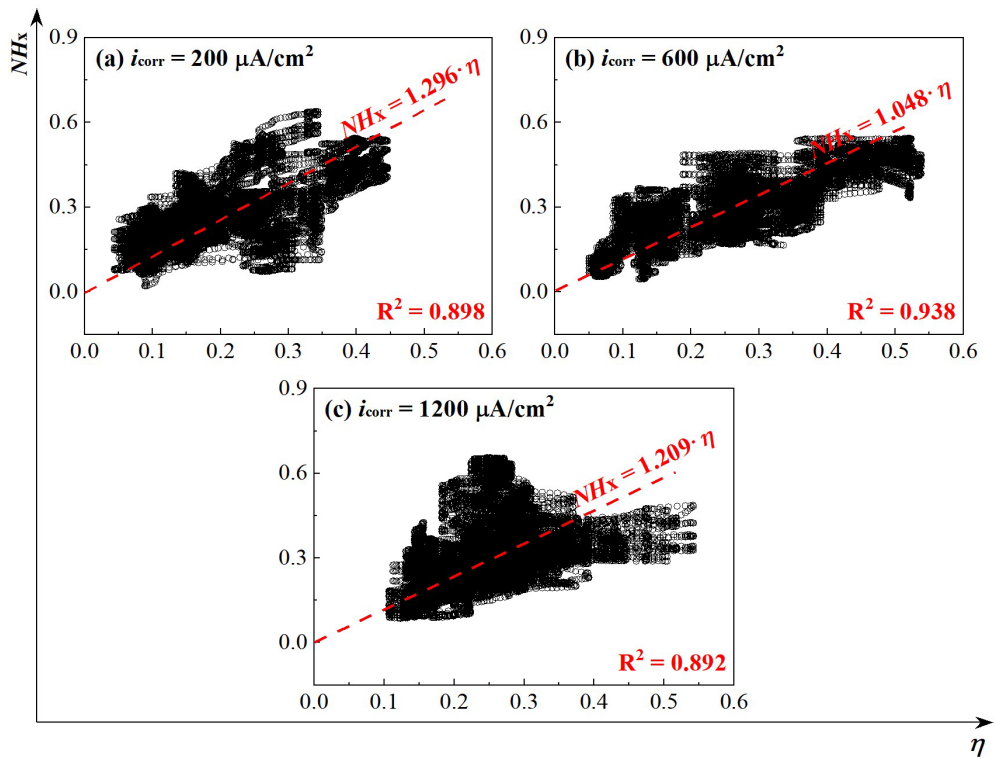

**Fig. 12.** The $NH_x$ sample points with corrosion current densities $i_{corr}$.

According to the actual distribution of $NH_x$ sample points in Fig. 12, the corrosion degree $\eta$ range of [0.04, 0.55] shown in Fig. 9 is divided into ten sub-intervals of [0.04, 0.10), [0.10, 0.15), [0.15, 0.20), [0.20, 0.25), [0.25, 0.30), [0.30, 0.35), [0.35, 0.40), [0.40, 0.45), [0.40, 0.45), [0.50, 0.55]. Subsequently, the Pearson correlation coefficients and JS divergence matrices across different $\eta$ subintervals of the $NH_x$'s PDD sequence combinations of different corrosion current densities $i_{corr}$ were calculated, as shown in Fig. 13. As the values of the $\eta$ subintervals increase, the mean Pearson correlation coefficient initially rises above 0.8 before falling below it, and the mean JS divergence first decreases to below 0.1 and then surpasses 0.1. The fundamental reason for this result is that the 8,592 $NH_x$ sample points within $\eta \in$ [0.04, 0.10) and the 4344 $NH_x$ sample points within $\eta \in$ [0.40, 0.55) account for only 8.52% and 4.31% of the total 100,800 $NH_x$ sample points, respectively. These relatively sparse sample points lead to greater random deviations in the $NH_x$'s PDDs in these two $\eta$ subintervals. In contrast, the $NH_x$ sample points within $\eta \in$ [0.10, 0.40), which make up 87.17% of the total, exhibit mean Pearson correlation coefficient values greater than 0.85 and mean JS divergence values of the $NH_x$'s PDD sequence combinations less than 0.085 across different $i_{corr}$ values.

Overall, the PDDs of $NH_x$ under varying corrosion current densities $i_{corr}$ also exhibit high similarity. Therefore, to simplify subsequent analyses, it is posited that the $i_{corr}$ does not affect the PDD characteristics of $NH_x$, with the existing differences stemming from random factors in the experiment.

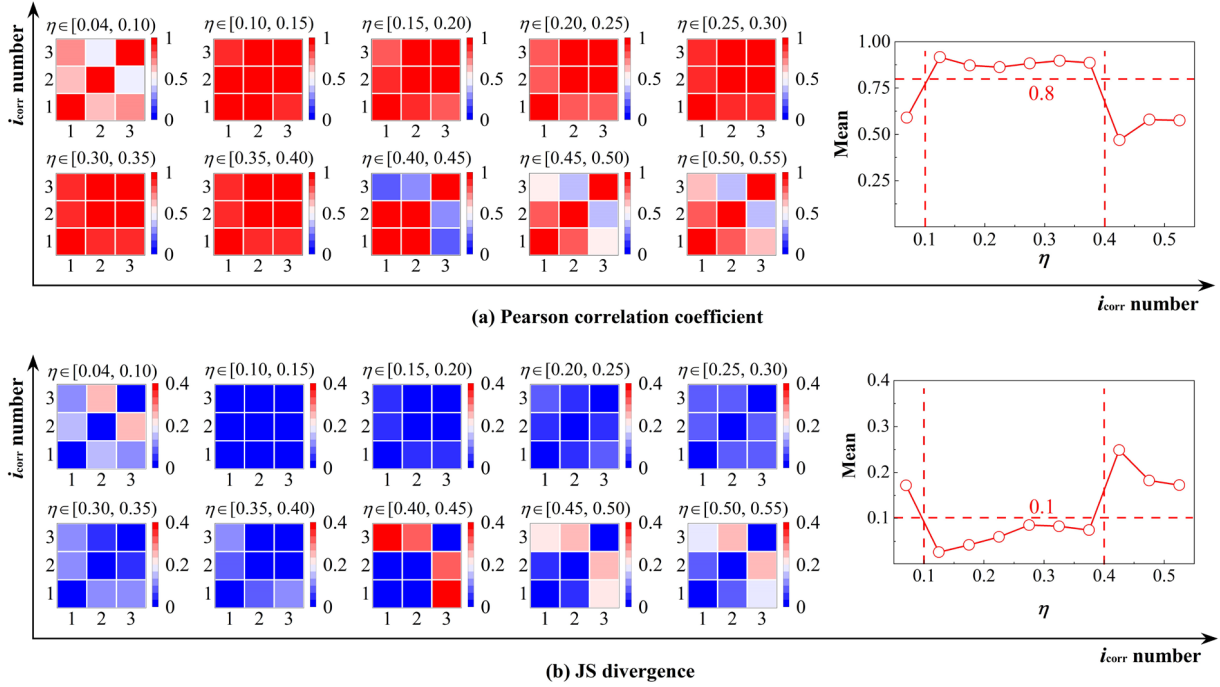

**Fig. 13.** (a) Pearson correlation coefficient and (b) JS divergence of PDD sequence combination of $NH_x$ at different corrosion current densities $i_{corr}$.

### 3.4.3. The probability density function (PDF) of index $NH_x$

According to the above analysis, it is believed that the $NH_x$ sample points of different LFHs and corrosion current densities $i_{corr}$ obey the same PDD. Therefore, in this section, all the $NH_x$ sample points shown in Fig. 8(b) are statistically analyzed to evaluate their overall PDD. The PDD estimation of $NH_x$ sample points is still conducted using the KDE method [10, 40].

Firstly, according to the actual distribution of $NH_x$ sample points in Fig. 8(b), the corrosion degree $\eta$ range of [0.04, 0.55] is divided into twelve sub-intervals of [0.04, 0.08), [0.08, 0.12), [0.12, 0.16), [0.16, 0.20), [0.20, 0.24), [0.24, 0.28), [0.28, 0.32), [0.32, 0.36), [0.36, 0.40), [0.40, 0.44), [0.44, 0.48), [0.48, 0.55]. The sub-interval [0.48, 0.55] is selected because the $NH_x$ sample points in this edge interval are too sparse and should not be further subdivided.

Then, the KDE method is employed to calculate the true PDD of $NH_x$ sample points within those above 12 $\eta$ sub-intervals, which aligns well with the sample distribution shown in the histogram, as illustrated in Fig. 14. The K-S test method [44] is used to examine the PDD type of $NH_x$ sample points, and the results indicate that the PDD of $NH_x$ sample points within each $\eta$ sub-interval follows the Weibull distribution function given by Eq. (7).

$$W(NH_x; \lambda, k) = \frac{k}{\lambda} \cdot \left(\frac{NH_x}{\lambda}\right)^{k-1} \cdot \exp\left(-\left(\frac{NH_x}{\lambda}\right)^k\right) \tag{7}$$

where $\lambda$ and $k$ are the scale parameter and shape parameter of the Weibull distribution function, respectively.

Subsequently, the Weibull function is employed to fit the PDD of $NH_x$ sample points within the 12 $\eta$ sub-intervals, yielding a coefficient of determination $R^2$ ranging from 0.899 to 0.999. This result indicates that the Weibull distribution function can accurately describe the true PDD of index $NH_x$.

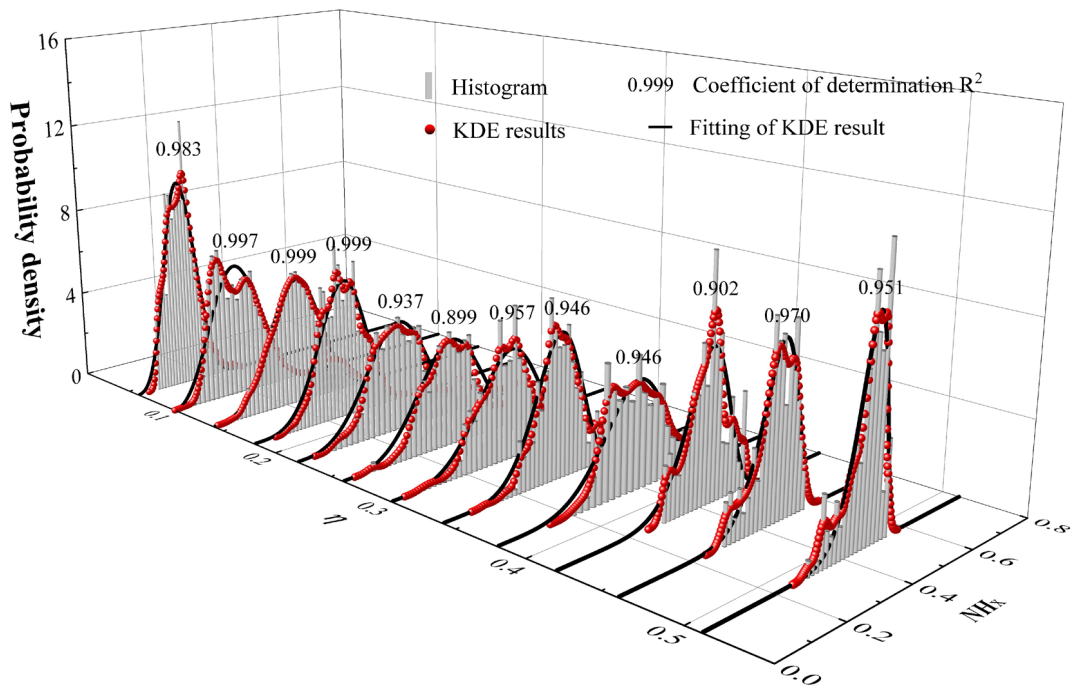

**Fig. 14.** The probability density of the index $NH_x$ of different $\eta$ sub-intervals.

Finally, the median values of $\eta$ for each subinterval are plotted on the X-axis, with the scale parameter $\lambda$ and shape parameter $k$ obtained from the data fitting using the Weibull distribution function, as shown in Fig. 15. Parameters $\lambda$ and $k$ increase linearly and powerly with the increasing $\eta$, respectively, yielding the coefficient of determination $R^2$ of 0.969 and 0.967, indicating high accuracy.

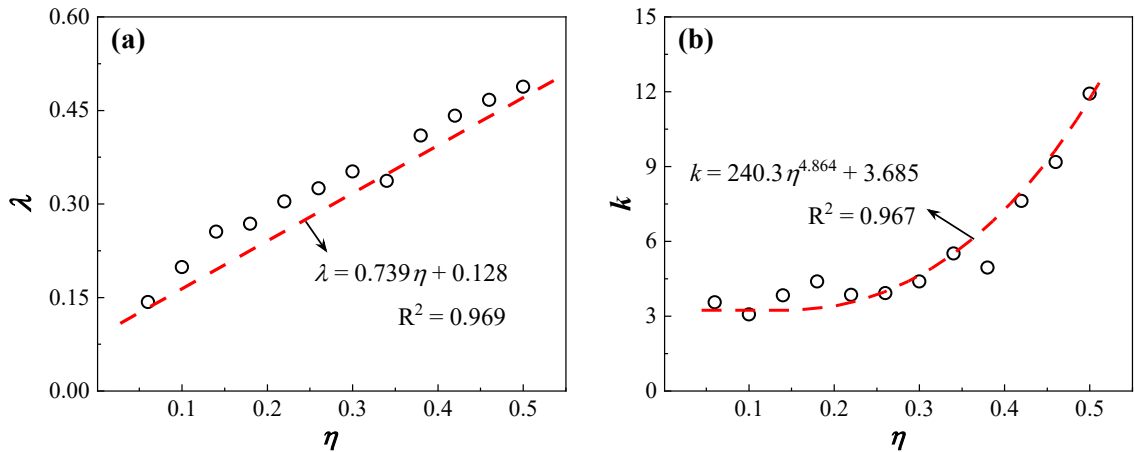

**Fig. 15.** The fitting results of the (a) scale parameter $\lambda$ and (b) shape parameter $k$ of the PDF of index $NH_x$.

Consequently, the probability density function (PDF) of index $NH_x$ can be expressed as follows:

$$\begin{cases} \text{PDF}(NH_x; \lambda, k) = \frac{k}{\lambda} \cdot \left(\frac{NH_x}{\lambda}\right)^{k-1} \cdot \exp\left(-\left(\frac{NH_x}{\lambda}\right)^k\right) \\ \lambda = 0.739 \cdot \eta + 0.128 \\ k = 240.3 \cdot \eta^{4.864} + 3.685 \end{cases} \quad (8)$$

The PDF of index $NH_x$ given in Eq. (8) results from the two fits, and its conformity degree with the real PDD of index $NH_x$ calculated by the KDE method needs to be clarified. Fig. 16 presents the PDD of index $NH_x$ calculated by the KDE method and Eq. (8) for the three-dimensional and two-dimensional forms, respectively. It can be observed that the PDD of index $NH_x$ calculated by Eq. (8) exhibits good consistency in both distribution shape and numerical values with the real PDD of index $NH_x$ calculated by the KDE method, indicating that Eq. (8) can effectively describe the PDD of $NH_x$.

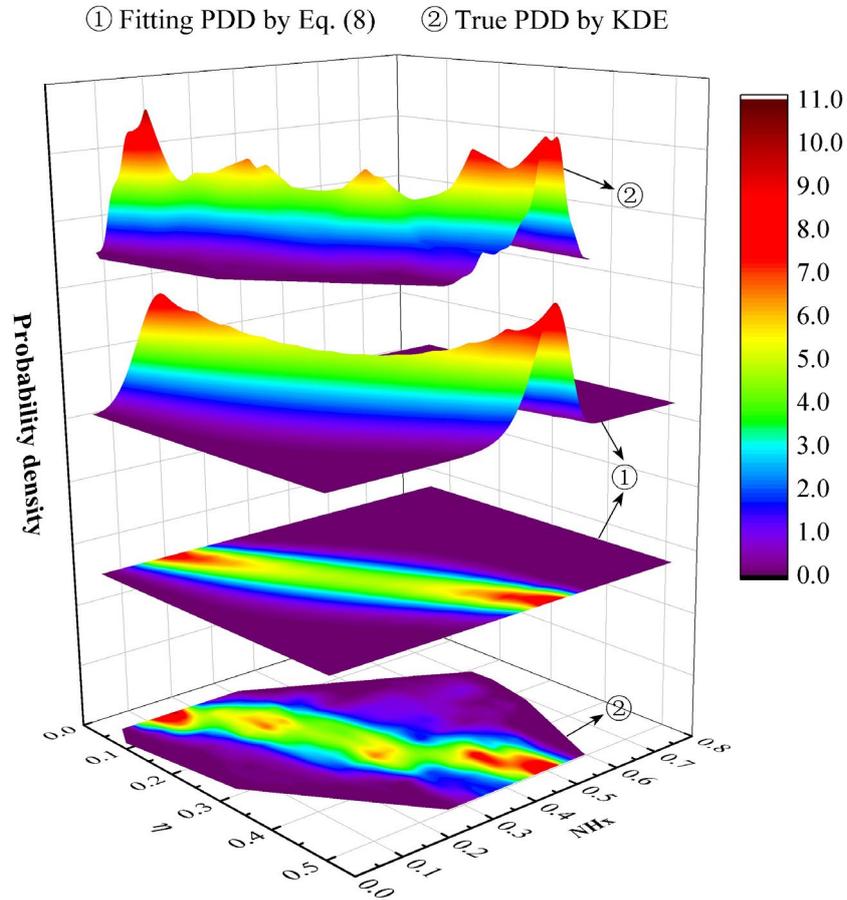

**Fig. 16.** The true and fitted PDD of index $NH_x$.

### 3.5. The Correction of the index $NH_x$

The PDF of the index $NH_x$, as given by Eq. (8), is established based on experimental data from rebar specimens S1 – S21, which have specific geometric dimensions. When utilizing the index $NH_x$ to estimate the corrosion degree $\eta$ of rebar with different geometric dimensions, it is necessary to consider the effect of changes in geometric dimensions on the index $NH_x$.

The quantitative index $NH_x$ follows the definition of a class of quantitative SMFL indices $I$, as outlined in reference [32]:

$$I = d \cdot f(x, y, z, w, l, r) \tag{9}$$

where $d$ represents the corrosion depth, $x$, $y$, and $z$ are the coordinate variables in a spatial Cartesian coordinate system, $w$ is the half-corrosion width, $l$ is the half-length of rebar, $r$ is the radius of rebar, and $f(\cdot)$ represents a function.

According to Eq. (9), the index $NH_x$ is positively correlated with $d$ while affected by the geometric variables $d$, $w$, $l$, and $r$, as well as the coordinate variables $x$, $y$, and $z$. In this study, the SMFL scanning path is positioned directly above the rebar, with the coordinate variable $y = 0$ being a constant [32]. In Fig. 6, the amplitude of $H_{Sx}$ with the same corrosion degree $\eta$ does not show significant variation with changes in the $x$ coordinate, except near the ends of the rebar or the corrosion boundaries. Therefore, the effect of the $x$ coordinate variable is considered to be included in the random error present in Eq. (8). Strictly speaking, the concept of corrosion width $2w$ does not apply in the case of entire corrosion and is, therefore, not taken into account. Consequently, the effects of the three variables $z$, $l$, and $r$ on the index $NH_x$ still need to be considered.

According to the concept of "scaling invariance" proposed by the author [33], proportional scaling of the rebar radius $r$, rebar half-length $l$, and the coordinate $z$ (that is, LFH) do not affect the values of the index $NH_x$ and the corrosion degree $\eta$. In Section 3.4.1, it has been demonstrated that the effect of LFH on the PDD of index $NH_x$ is minimal, and thus, it is believed that the rebar scaling-induced variations in $r$ and $z$ do not influence the correlation between the index $NH_x$ and the corrosion degree $\eta$. Consequently, only the effect of the rebar half-length $l$ on the index $NH_x$ needs to be considered.

The correction factor $CR$ is introduced to measure the effect of the rebar half-length $l$ on

the index $NH_x$:

$$CR=\frac{NH_x(2l)}{NH_x(1.5)} \quad (10)$$

where $NH_x(1.5)$ is the theoretical $NH_x$ value at the midpoint of the standard corroded rebar with $r = 7$ mm, $2l = 1.5$ m, and LFH = 30 mm, and $NH_x(2l)$ is the theoretical $NH_x$ value at the midpoint of a corroded rebar with $r = 7$ mm, a length of $2l$, and LFH = 30 mm.

The calculation data range for $H_{\text{I-av}}$ in Eq. (3) can be determined as $x \in [-5l/6, 5l/6]$. The theoretical value of the correction factor $CR$ can be obtained by utilizing the analytic expressions for $H_{\text{I}x}$ and $H_{\text{I}z}$ provided in reference [33] and Eqs. (3) and (10). Alternatively, the numerical solution for $CR$ can be derived using the finite element simulation method detailed in reference [5]. Theoretically, when $\eta$ remains constant, the correction factor $CR$ represents the proportional change in the index $NH_x$ due to variations in the length of the standard corroded rebar. Therefore, for a corroded rebar with a radius of $r$ and a length of $2l$, it should first be scaled to a rebar with a radius of 7 mm and a length of $(7/r)\cdot 2l$. Subsequently, the correction factor $CR$ can be calculated based on the scaled length of $(7/r)\cdot 2l$. Finally, the corrected index $NH_{\text{x-m}}$ can be expressed as:

$$NH_{\text{x-m}}=\frac{NH_x}{CR} \quad (11)$$

where $NH_x$ is the measured value of the corroded rebar with radius $r$ and length $2l$.

Using corrected index $NH_{\text{x-m}}$, the effect of the rebar half-length $l$ on the estimation of corrosion degree $\eta$ based on $NH_x$ and Eq. (8) can be eliminated.

## 4. Index $NH_x$-based quantitative estimation of the corrosion degree $\eta$

As shown in Fig. 16, for a given $NH_x$, the corresponding $\eta$ exhibits a certain level of randomness, which makes it impossible to directly determine $\eta$ using $NH_{x\text{-}m}$ and Eq. (8). To address this issue, a specific naive Bayesian model as expressed in the following is employed:

$$\pi(\eta|NH_x) = \frac{f(NH_x|\eta) \cdot \pi(\eta)}{\int_\Theta f(NH_x|\eta) \cdot \pi(\eta) d(\eta)} \tag{12}$$

where $\pi(\eta|NH_x)$ is the posterior distribution, $f(NH_x|\eta)$ is the likelihood function, $\pi(\eta)$ is the prior distribution, and $\Theta$ represents the parameter space (in this study, $\Theta$ is the range of $\eta$).

Taking specimen S1 as an example, a probabilistic estimation of the corrosion degree $\eta$ of specimen S1 based on this Bayesian model and the measured $NH_x$ sequences is conducted. Eq. (8) is used as the likelihood function $f(NH_x|\eta)$, and to obtain the posterior distribution $\pi(\eta|NH_x)$, the prior distribution $\pi(\eta)$ needs to be determined. Statistical analysis of the measured 801 corrosion degree $\eta$ samples of specimen S1 within the corrosion area of 350 – 1150 mm yields the PDD of the true corrosion degree $\eta$, as shown in Fig. 17. The true PDD of $\eta$ approximates a Weibull distribution, and thus the fitted Weibull PDF $W(\eta; 0.35, 4.7)$ is used as the informative prior distribution $\pi(\eta)$. The non-informative prior distribution is represented by a uniform distribution $U(1.0)$ with a constant probability density of 1.0. Additionally, for comparative analysis, a prior distribution $W(\eta; 0.1, 1.5)$, which significantly deviates from the true PDD of $\eta$, is also provided.

Based on Eq. (8), Eq. (12), Fig. 17, and the measured 801 $NH_x$ values, the posterior distribution of the corrosion degree $\eta$ at different cross-sections along the longitudinal axis of specimen S1 can be obtained, as shown in Fig. 18. It can be observed that the corrosion degree $\eta$ estimation results derived from the prior distributions W($\eta$; 0.35, 4.7) and U(1.0) are close to the true values, whereas the estimation result derived from the prior distribution W($\eta$; 0.1, 1.5) deviate significantly to the true values. Therefore, to obtain the most accurate estimation result of the corrosion degree $\eta$, it is essential to select an appropriate prior distribution based on the specific conditions encountered in practice. Using the probability estimation result of the corrosion degree $\eta$, it is possible to analyze and derive corrosion degree $\eta$ estimates of the corroded rebar at a certain level of confidence.

Therefore, once the measured value of the SMFL quantitative index $NH_x$ of any entirely corroded rebar is obtained non-destructively, its cross-sectional corrosion degree $\eta$ can be automatically estimated from a probabilistic perspective.

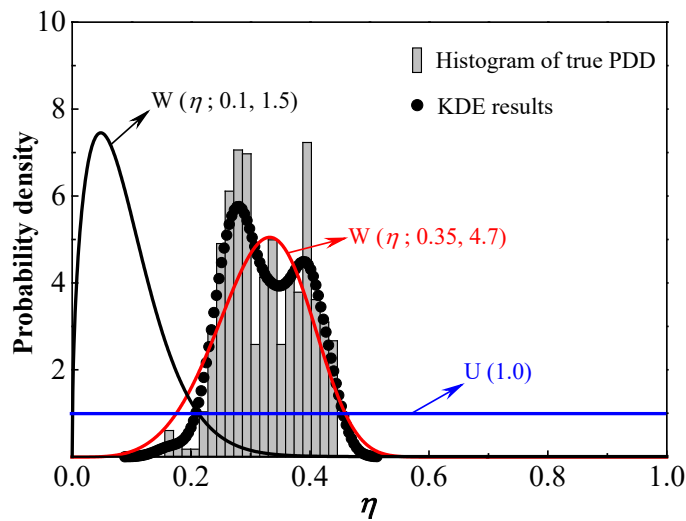

**Fig. 17.** Three corrosion degree prior distributions $\pi(\eta)$ of specimen S1.

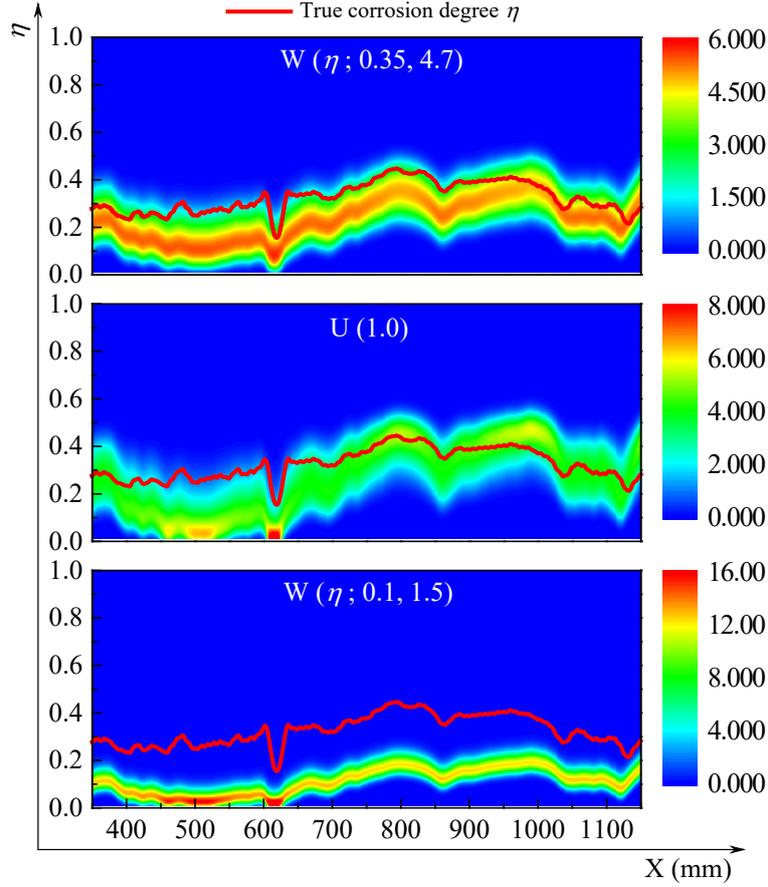

**Fig. 18.** The posterior distribution of corrosion degree $\eta$ of different cross-sections of specimen S1 under different prior distributions $\pi(\eta)$.

# 5. Conclusions

In this study, the correlation between the self-magnetic flux leakage (SMFL) variation and corrosion degree of the entirely corroded rebars was established using a large amount of high-precision magnetic field and 3D scanning test data, based on which a novel non-destructive quantitative probabilistic estimation method of the cross-sectional corrosion degree of rebar was proposed. The following conclusions can be drawn:

(1) The corrosion of rebars causes significant variations in the SMFL field. The location and range of the rebar corrosion can be accurately determined by the SMFL field variations, with an error not exceeding 3.0%.

(2) The proposed SMFL quantification index $NH_x$ effectively mitigates the adverse effects of differences in magnetization states and exhibits a significant linear correlation with the rebar cross-sectional corrosion degree $\eta$. The probability density distribution of $NH_x$ can be adequately described by the Weibull distribution function, with its distribution parameters $\lambda$ and $k$ demonstrating strong linear and power relationships with $\eta$, respectively.

(3) Based on the prior distribution of $\eta$, the probability density function of $NH_x$, and a Bayesian model, the rebar's accurate $\eta$ corresponding to any specific non-destructively measured $NH_x$ value can be automatically estimated. Hence, the SMFL-based nondestructive quantitative estimation of rebar cross-sectional corrosion degree is achieved.

However, the proposed probability density function of $NH_x$ is based on limited SMFL information, which is not ideal. Future work could incorporate machine learning methods to more fully utilize the SMFL information, thereby yielding more accurate assessment results.

## Acknowledgement

The authors gratefully acknowledge the financial support from the Program of Shanghai Science and Technology Committee (22dz1203603) and the National Natural Science Foundation of China (51878486). The authors also sincerely thank the anonymous reviewers for their thorough reviews and constructive comments and the editors for their selfless contributions in the manuscript processing.

## Data availability statement

The raw/processed data required to reproduce these findings will be shared by the corresponding author upon reasonable request.

predictive models. ACI Struct J 2005; 102(5): 719.

[38] Yuan Y, Ji Y, Shah SP. Comparison of two accelerated corrosion techniques for concrete structures. ACI Struct J 2007; 104(3): 344.

[39] Geng J. Structured-light 3D surface imaging: a tutorial. Ad Opt Photonics 2011; 3(2): 128-160.

[40] Parzen E. On estimation of a probability density function and mode. Ann Math Stat 1962; 33(3): 1065-76.

[41] Li Z, Gao X, Lu D. Correlation analysis and statistical assessment of early hydration characteristics and compressive strength for multi-composite cement paste. Constr Build Mater 2021; 310: 125260.

[42] Chu J, Xiao X. Benefits evaluation of the northeast passage based on grey relational degree of discrete Z-numbers. Inform Sciences 2023; 626: 607-625.

[43] Li R, Qian Fl, Du Xq, Zhao S, Zhang YP. A collaborative filtering recommendation framework based on Wasserstein GAN. J Phys Conf Ser 2020; 1684(1): 012057.

[44] Conover WJ. Practical nonparametric statistics. John Wiley & Sons, 1999.